\documentclass[12pt,a4paper]{article}
\usepackage{amsmath}
\usepackage{amssymb}
\usepackage{enumerate}
\usepackage{amsfonts}
\usepackage[latin1]{inputenc}
\begin{document}

%
\newcounter{saveeqn}
\newcommand{\alpheqn}{\setcounter{saveeqn}{\value{equation}}%
\setcounter{equation}{0}%
\addtocounter{saveeqn}{1}
\renewcommand{\theequation}{\mbox{\arabic{saveeqn}\alph{equation}}}}
\newcommand{\reseteqn}{\setcounter{equation}{\value{saveeqn}}%
\renewcommand{\theequation}{\arabic{equation}}}
\begin{center}
\renewcommand{\thefootnote}{*}

{\bf `` D-BRANE ACTIONS AS CONSTRAINED SYSTEMS'' }
\\[14mm]
Usha Kulshreshtha$^{[{\rm a}]}$ and D.S. Kulshreshtha\footnote{ ``Plenary Talk'' presented by DSK at THEP-I,  I.I.T. Roorkee, Roorkee - 247667, India, March 16-20, 2005 (To appear in the Conference Proceedings).}$^{[{\rm b}]}$ 
\end{center}

\begin{tabular}{cc}
a & Department of Physics, Swami Shrdhanand College, \\
& University of Delhi, Delhi, India. \\
& $<$ushakulsh@netscape.net$>$ \\
b & Department of Physics and Astrophysics,\\
&  University of Delhi, Delhi, India. \\
& $<$dsk@physics.du.ac.in$>$ 
\end{tabular} 

\vspace{5cm}

\begin{abstract}
After a brief introduction to D-brane actions, the constrained dynamics and the constraint quantization of some D-brane actions is considered.
\end{abstract}

\newpage
\section{Introduction}

The main aim of science has been to find an unified picture of nature. The standard model of particle physics which describes the quantization of the electromagnetic field and the strong and weak nuclear forces is one of the most accurately tested thoeries of nature. However, it does not contain gravity. String theory (ST)\cite{1} is the best candidate for a theory of everything (TOE). It  tries to unify all the four fundamental forces of nature in to a singele unified framework. In ST, the different particles are different vibrational states of a string whose typical size is of the order of Planck length and they appear to be point like to the present day accelerators. In ST, particles and fields are replaced by one dimensional strings that propagate in an ambient spacetime. Also, the point vertex interaction of quantum field theory involving three world lines, in ST, is replaced by a more spreaded out vertex of ST, involving world sheets (WS) of either open or closed strings. String interactions therefore do not occure at one point (as it happens in field theory) and it also leads to a more sensible quantum behaviour.

The strings open or closed when move through spacetime sweep out an imaginary surface called a WS, and higher dimensional objects called D-branes swep out the imaginary volume called world volume (WV). The strings can have various kinds of boundary conditions (BC's): closed strings e.g., have periodic BC's where as open strings have two different types of BC's called Neuman BC's and Dirichlet BC's. With Neuman BC's, the end point is free to move about but no moment flows out. With the Dirichlet BC's, however, the end point is fixed to move only on some manifold which is called a D-brane or Dp-brane, where p is an integer and is equal to the number of spatial dimensions of the manifold. D-branes are dynamical objects (hypersurfaces embeded in spacetime) have fluctuations and can move around. In the popular terminology, a D0-brane is a point particle, a D1-brane is a string, a D2-brane is a membrane, and so on\cite{1}.

Quite often the ST's like most of the field theories are the constrained systems \cite{2,3,4} that are defined in terms of overdetermined set of coordinates and need to be quantized using the conventional methods of constraint quantization \cite{2,3,4} in terms of the various Dirac's relativistic forms of dynamics\cite{5}. In our work we use ,in particular, the so-called the instant-form (IF)\cite{5} and front-form (FF)(or the light-cone (LC) quantization (LCQ))\cite{5}. In the following, we would consider the constrained dynamics of some D-brane actions using the IF and FF/LCQ. As examples, we would consider the Nambu-Goto action (NGA), Born-Infeld-Nambu-Goto action (BINGA), Dirac-Born-Infeld-Nambu-Goto action (DBINGA), Polyakov action (PA), and the conformally gauge-fixed Polyakov D1-brane action (CGFPD1BA) with the two-form gauge field $B_{\alpha\beta}$ and discuss their constrained dynamics and quantization \cite{1,2,3,4}.

\section{\bf The Nambu-Goto Type Actions}
\subsection{\bf Nambu-Goto Action}
The Nambu-Goto type actions are the ones that involve a square root.
The NGA describing the propagation of a D1-brane in a $d$-dimensional flat background (with $d = 10$ for the fermionic and $d = 26$ for bosonic case) is defined by \cite{1,2}:

\alpheqn
\begin{eqnarray}
S_1 &=& \int {\cal L}_1~~ d^2\sigma \\
\label{1a}
{\cal L}_1 &=& [-T]\left[ {[-det(G_{\alpha\beta})]}^\frac {1}{2} \right] \\
\label{1b}
G_{\alpha\beta} &=& \partial_\alpha X^\mu \partial_\beta X^\nu \eta_{\mu\nu} \quad ; \quad \alpha,\beta = 0,1 \\
\label{1c}
\eta_{\mu\nu} &=& {\rm diag}(-1,+1,...+1); \quad \mu,\nu = 0,1,2,...,(d-1). 
\label{1d}
\end{eqnarray}
\reseteqn
In the present work we would consider only the bosonic D1-brane (sometimes also called the D-string) with $d = 26$ (however, for the corresponding fermionic case one has $d = 10$ ). Here $\sigma^\alpha \equiv (\tau, \sigma)$ are the two parameters describing the WS. In the following, by overdots and primes, we would denote in general, the derivatives with respect to the WS coordinates $\tau$ and $\sigma$. The string tension $T$ is a constant of mass dimension two. $G_{\alpha\beta}$ is the induced metric on the WS and $X^\mu(\tau, \sigma)$ are the maps of the WS into the $d$-dimensional Minkowski space and describe the strings evolution in space time \cite{1,2}.

 The theory described by $S_1$ is seen to possess two primary constraints \cite {2,3,4}:
\begin{eqnarray}
\Psi_1 = (\Pi.X^{'}) \approx 0 \quad ; \quad \Psi_2 = [\Pi^2 + T^2 (X^{'})^2] \approx 0
\label{2}
\end{eqnarray}
Where $\Pi^\mu$ are the canonical momenta conjugate respectively to $X_\mu$. Here the symbol $ \approx $  denotes a weak equality (WE) in the sense of Dirac \cite {2,3,4}, and it implies that these above constraints hold as strong equalities only on the reduced hypersurface of the constraints and not in the rest of the  phase space of the classical theory (and similarly one can consider it as a weak operator equality (WOE) for the corresponding quantum theory). The above constraints are infact, the usual so-called Virasoro constraints of the theory \cite {1}.

The canonical Hamiltonian density corresponding to ${\cal L}_1$ is seen to vanish identically and  therefore the dynamics of the system is completely determined by the constraints of the theory and the total Hamiltonian density of the theory which is obtained by incorporating the primary constraints of the theory in the canonical Hamiltonian density with the help of Lagrange multiplier fields,  which are to be treated as dynamical. The Hamilton's equations of motion obtained from the total Hamiltonian preserve the constraints of the theory in the course of time. Also the theory is seen to possess only the above two constraints which form a set of first-class constraints and the theory is indeed seen to be gauge-invariant (GI) (and consequently gauge nonanomalous) and it possesses the usual three local gauge symmetries given by the two-dimensional WS reparametrization invariance (WSRI) and the Weyl invariance (WI) \cite{1,2}, and it could be quantized under suitable gauge-fixing conditions(GFC's).

\subsection{Born-Infeld-Nambu-Goto Action (BINGA)}

The BINGA describing the propagation of a D1-brane  in a $ d $ - dimensional flat background (with $d = 10$ for the fermionic and $d = 26$ for the bosonic case) is defined by \cite{1,2}:

\alpheqn
\begin{eqnarray}
S_2 &=& \int {\cal L}_2~ d^2 \sigma \quad ; \quad
{\cal L}_2 = (-T) [-\det (G_{\alpha\beta} + F_{\alpha\beta})]^{\frac12} \\
\label{3a}
G_{\alpha\beta} &=& \partial_\alpha X^\mu \partial_\beta X^\nu \eta_{\mu\nu};
 ~F_{\alpha\beta} = (\partial_\alpha A_\beta - \partial_\beta A_\alpha) ; \quad \alpha,\beta = 0,1  \\
\label{3b}
\eta_{\mu\nu} &=& {\rm diag}(-1,+1,...+1); \quad \mu,\nu = 0,1,2,...,(d-1) ; \quad f = F_{01} = -F_{10} 
\label{3c}
\end{eqnarray}
\reseteqn
In the present work we would consider only the bosonic D1-brane with $d = 26$. Here $F_{\alpha\beta}$ is the Maxwell field strength of the $ U(1)$ gauge field  $A_{\alpha} (\tau, \sigma) $. It is important to mention here that the $ U(1)$ gauge field $ A_{\alpha} $ is a scalar field in the target-space whereas it is an $ {\alpha}$ - vector field in the WS-space. Also, we are considering  the $U(1)$ gauge field $A_{\alpha}$ to be a function only of the WS coordinates $\tau$ and $ \sigma $ and not of the target-space coordinates $ X^\mu$. The theory described by $S_2$ is seen to possess three primary constraints:
\begin{eqnarray}
\psi_1 = \pi^0 \approx 0 \quad ; \quad
\psi_2 = (\Pi \cdot X') \approx 0 \quad ; \quad
\psi_3 = [\Pi^2 + (E^2+ T^2)(X')^2] \approx 0
\label{4}
\end{eqnarray}
Where $\Pi^\mu, \pi^0$ and $E(\equiv \pi^1)$ are the canonical momenta conjugate respectively to $X_\mu, A_0$ and $A_1$. Demanding that the primary constraint $\psi_1$ be preserved in the course of time one obtains a secondary constraint:
 $ \psi_4 $ =  $ [E^{'}] \approx 0 $. The preservation of $\psi_4$ for all time does not give rise to any further constraints. Similarly, the preservation of $\psi_2$ and $\psi_3$ for all time also does not yield any further constraints.
The theory is thus seen to possess only four constraints $\psi_1, ~ \psi_2,~\psi_3$, and $~\psi_4$, which form a set of first-class constraints. The theory is indeed seen to be GI possessing the usual local gauge symmetries given by the WSRI and the WI, and it could be quantized under appropriate gauge-fixing.

\subsection{Dirac-Born-Infeld-Nambu-Goto Action (DBINGA)}

The DBINGA  describing the propagation of a D1-brane in a $d$-dimensional flat background (with $d = 10$ for the fermionic and $d = 26$ for the bosonic case) is defined by \cite{1,2}:

\alpheqn
\begin{eqnarray}
S_3 &=& \int {\cal L}_3~ d^2 \sigma   \\
\label{5a}
{\cal L}_3 &=& (-T) [-\det (G_{\alpha\beta} + {\cal F}_{\alpha\beta}]^{\frac12} \\
\label{5b}
G_{\alpha\beta} &=& \partial_\alpha X^\mu \partial_\beta X^\nu \eta_{\mu\nu} \\
\label{5c}
{\cal F}_{\alpha\beta} &=& (F_{\alpha\beta} - B_{\alpha\beta}); \quad F_{\alpha\beta} = (\partial_\alpha A_\beta - \partial_\beta A_\alpha) ; \quad \alpha, \beta = 0,1 \\
\label{5d}
B_{\alpha\beta}& := &\partial_{\alpha}X^{\mu} \partial_{\beta}X^{\nu}B_{\mu\nu} ; \quad  B_{\alpha\beta} = \left( \begin{array}{ll} ~~0 & b \\ -b & 0 \end{array} \right) \quad ; \quad  b = B_{01} = - B_{10} \\
\label{5e}
\eta_{\mu\nu} &=& {\rm diag}(-1,+1,...+1); \quad \mu,\nu = 0,1,2,...,(d-1) 
\label{5f}
\end{eqnarray}
\reseteqn
In the present work we would consider only the bosonic D1-brane with $d = 26$. Here $F_{\alpha\beta}$ is the Maxwell field strength of the U(1) gauge field $A_\alpha (\tau, \sigma), $ and $B_{\alpha\beta} (\tau, \sigma)$ is a constant background antisymmetric 2-form gauge field. It is important to mention here that the 2-form gauge field $ B_{\alpha\beta} $ is a scalar field in the target-space whereas it is an antisymmetric tensor field in the WS-space.Also, we are considering the 2-form gauge field $B_{\alpha\beta} $ as well as the $U(1)$ gauge fields $A_{\alpha}$, to be functions only of the WS coordinates $\tau$ and $ \sigma $ and not of the target-space coordinates $ X^\mu$. The theory is seen to possess four primary constraints\cite{2}:
\begin{eqnarray}
\psi_1 = \Pi_b \approx 0 ; \quad \psi_2  = \pi^0 \approx 0 ; \quad \psi_3  = (\Pi \cdot X') \approx 0 ;\quad \psi_4 = [\Pi^2 + (E^2+ T^2)(X')^2] \approx 0
\label{6}
\end{eqnarray}
Where $\Pi^\mu, \pi^0, E(\equiv \pi^1)$ and $\Pi_b$ are the canonical momenta conjugate respectively to $X_\mu, A_0, A_1$ and $b(=B_{01} = - B_{10})$. Demanding that the primary constraint $\psi_1$ be preserved in the course of time one obtains a secondary constraint: $\tilde \psi_5 $ = $ [-E] \approx 0 $.
The preservation of $\psi_2$ for all time gives rise to another secondary constraint: $ \tilde \psi_6$ = $(E') \approx 0 $. The preservation of $\tilde \psi_5$ and $\tilde \psi_6$ for all time does not give rise to any further constraints and similarly the preservation of $\psi_3$ and $\psi_4$ for all time also does not yield any further constraints. Further the constraint $\tilde \psi_5$ also implies the constraint $\tilde \psi_6$ and therefore it is enough to consider only one constraint namely, $\psi_{5}$ = $ E$ $ \approx 0$.  In view of this, one could consider only one secondary constraint namely, $\psi_5$ (and not $\tilde \psi_5$ and $\tilde \psi_6$). The theory is thus seen to possess only five constraints $\psi_1, ~ \psi_2,~\psi_3,~\psi_4$ and $\psi_5$, which form a set of first-class constraints. The theory is indeed seen to be GI possessing the usual local gauge symmetries given by the WSRI and the WI and could be quantized under suitable GFC's\cite{2}.

\section{ Polyakov Action}

The Polyakov action describing the propagation of a D1-brane in a $d$-dimensional curved background $h_{\alpha \beta}$ is defined by \cite{1,2}:
\alpheqn
\begin{eqnarray}
\tilde S &=& \int \tilde{\cal L} d^2\sigma \quad ; \quad
\tilde{\cal L} = \left[ - \frac T2 \sqrt{-h} h^{\alpha\beta} G_{\alpha\beta} \right]; \quad h = \det (h_{\alpha\beta}) \\
\label{7a}
G_{\alpha\beta} &=& \partial_\alpha X^\mu \partial_\beta X^\nu \eta_{\mu\nu}; ~ \eta_{\mu\nu} = {\rm diag} (-1,+1,.....,+1) \\
\label{7b} 
\mu,\nu &=& 0,1,........,(d-1); ~ \alpha,\beta = 0,1
\label{7c} 
\end{eqnarray}
\reseteqn
Here  $h_{\alpha\beta}$ are the auxiliary fields (which turn out to be proportional to the metric tensor $\eta_{\alpha\beta}$ of the two-dimensional surface swept out by the string). One can think of $\tilde S$ as the action describing $d$ massless scalar fields $X^\mu$ in two dimensions moving on a curved background $h_{\alpha\beta}$. Also because the metric components $h_{\alpha\beta}$ are varied in the above equation, the 2-dimensional gravitational field $h_{\alpha\beta}$ is treated not as a given background field, but rather as an adjustable quantity coupled to the scalar fields \cite{2}. The action $\tilde S$ has the well-known three local gauge symmetries described by the 2-dimensional WSRI and the WI given by \cite{1,2}:
\alpheqn
\begin{eqnarray}
X^\mu &\longrightarrow & \tilde X^\mu = [X^\mu + \delta X^\mu] \quad ; \quad 
\delta X^\mu = [\zeta^\alpha (\partial_\alpha X^\mu)] \\
\label{8a}
h^{\alpha\beta} &\longrightarrow& \tilde{h}^{\alpha\beta} = [h^{\alpha\beta} + \delta h^{\alpha\beta}] \quad ; \quad
\delta h^{\alpha\beta} = \left [\zeta^{\gamma} \partial_\gamma h^{\alpha\beta} - \partial_\gamma \zeta^\alpha h^{\gamma\beta} - \partial_\gamma \zeta^\beta h^{\alpha\gamma} \right ] \\
\label{8b}
h_{\alpha\beta} &\longrightarrow & [ \Omega  h_{\alpha\beta}]  
\label{8c}
\end{eqnarray}
\reseteqn
Where the WSRI is defined for the two parameters $\zeta^\alpha \equiv \zeta^\alpha(\tau,\sigma)$, and the WI is specified by a function $\Omega \equiv \Omega(\tau,\sigma)$ \cite{1,2}.

Now for studying the so-called CGFPD1BA one makes use of the fact that the 2-dimensional metric  $h_{\alpha\beta}$  is also specified by three independent functions as it is a symmetric ${2\times 2}$ metric. One can therefore use these gauge symmetries of the theory to choose $h_{\alpha\beta}$ to be of a particular form \cite{1,2}:    
\begin{eqnarray}
h_{\alpha\beta} := \eta_{\alpha\beta} \quad ; \quad h^{\alpha\beta} := \eta^{\alpha\beta}
\label{9}
\end{eqnarray}
which is called as the conformal gauge.  For studying the theory in the IF of dynamics (also called as the equal WS time  (EWST) framework) on the hyperplanes defined by the WS time $\tau = $ constant, we take \cite{1,2}:
\alpheqn
\begin{eqnarray}
h_{\alpha\beta} &:=& \eta_{\alpha\beta} = 
\left( \begin{array}{ll} -1 & ~~0 \\ ~~0 & +1 \end{array} \right) \quad ; \quad
h^{\alpha\beta} = \eta^{\alpha\beta} = 
\left( \begin{array}{ll} -1 & ~~0 \\ ~~0 & +1 \end{array} \right)\\
\label{10a}
\sqrt{-h} &=& \sqrt{-\det(h_{\alpha\beta})} = +1
\label{10b}
\end{eqnarray}
\reseteqn
In the FF/LCQ (also called as the equal LC-WS time (ELCWST) framework) of the theory we study the dynamics of the theory on the hyperplanes defined by the LC-WS time $\sigma^+ := (\tau+\sigma)=$ constant, and use the LC variables defined by \cite{1,2,5}:
\begin{equation}
\sigma^\pm := (\tau \pm \sigma) \quad {\rm and} \quad X^\pm := (X^0 \pm X^1)/\sqrt 2
\label{11}
\end{equation}
In this case we take :
\alpheqn
\begin{eqnarray}
h_{\alpha\beta} &:=& \eta_{\alpha\beta} = 
\left( \begin{array}{ll}  ~~~~0 & -1/2 \\ -1/2 & ~~~~0 \end{array} \right) \quad ; \quad
h^{\alpha\beta} := \eta^{\alpha\beta} = \left( \begin{array}{ll}  ~~0 & -2 \\ -2 & ~~0 \end{array} \right) \\
\label{12a}
\sqrt{-h} &=& \sqrt{- \det(h_{\alpha\beta)}} = + 1/2
\label{12b}
\end{eqnarray}
\reseteqn 
Now the action $\tilde {S} $ in the so called conformal-gauge finally reads \cite{1,2}:
\alpheqn
\begin{eqnarray}
S^{N} &=& \int {\cal L}^{N} d^2\sigma \quad ; \quad
{\cal L}^{N} = [(-T/2)] [ \partial^\beta X^\mu \partial_\beta X_\mu ] \\
\label{13a}
\beta &=& 0,1~~~~ and ~~~\mu = 0,1,i~; ~~~i = 2,3,...,25~~ (IF) \\
\label{13b}
\beta &=& +,-~~~~ and ~~~\mu = +,-,i~; ~~~i = 2,3,...,25~~ (LC/FF)
\label{13c}
\end{eqnarray}
\reseteqn
The action $S^{N}$ is the conformally gauge-fixed Polyakov D1-brane action (CGFPD1BA). This action is seen to lack the local gauge symmetries defined by Eqs.(8). This is in contrast to the fact that the original action $\tilde{S}$ had the local gauge symmetries defined by (8) and was therefore GI. The theory defined by the action  $S^{N}$, on the other hand describes GNI and consequently gauge anomalous theory. This is not surprising at all because the theory defined by $S^{N}$ is afterall a gauge-fixed theory and consequently not expected to be GI anyway. Infact, the theory defined by $S^{N}$ in the IF of dynamics does not possess any Dirac constraints, whereas in the FF/LCQ it is seen to possess a set of 26 second-class constraints \cite{2}: 
\alpheqn
\begin{eqnarray}
\Omega_1 &=& (P^+ + \frac T2 \partial_-X^+) \approx 0 \quad ; \quad
\Omega_2 = (P^- + \frac T2 \partial_-X^-) \approx 0 \\
\label{14a}
\Omega_3 &=& (P_i + \frac T2 \partial_-X^i) \approx 0 ; \quad i=2,3,.......,25 
\label{14b}
\end{eqnarray}
\reseteqn
Where  $P^+,~ P^-$ and $P_i$  are the momenta canonically conjugate respectively to $X^-,~ X^+$ and $X_i$. The theory described by $S^{N}$ in the IF as well as in the FF of dynamics is thus seen to be gauge anomalous and GNI and therefore does not possess the local gauge symmetries defined by WSRI and WI. We now consider this CGFPD1BA in the presence of a constant background antisymmetric NSNS 2-form gauge field.

\section{CGFPD1BA with a 2-form Gauge Field} 

The CGFPD1BA in the presence of a constant background antisymmetric 2-form gauge field $B_{\alpha\beta}$  is defined by \cite{1,2}:

\alpheqn
\begin{eqnarray}
S^{I} &=& \int {\cal L}^{I}~~ d^2\sigma \quad ; \quad
{\cal L}^{I} = [ {\cal L}^{C} + {\cal L}^{B} ] \\
\label{15a}
{\cal L}^{C} &=& [\lambda {\cal L}^{N}] = \left [- \frac{T}{2} \right ] [ \lambda \partial^{\beta} X^\mu \partial_\beta X_\mu ] \quad ; \quad
{\cal L}^{B} = \left [- \frac{T}{2} \right ] [\Lambda \varepsilon^{\alpha\beta} B_{\alpha\beta}] \\
\label{15b} 
\lambda &=& \sqrt{(1+\Lambda^2)};~~ \Lambda = constant ;  ~~\varepsilon^{\alpha\beta} = \left( \begin{array}{ll} ~~0 & 1 \\ -1 & 0 \end{array} \right ) \\ \label{15c}
B_{\alpha\beta}&:=& \partial_{\alpha} X^{\mu} \partial_{\beta}X^{\nu} B_{\mu\nu} \\ \label{15d}
B_{\alpha\beta} &=& \left( \begin{array}{ll} ~~0 & b \\ -b & 0 \end{array} \right);~~b = B_{01} = -B_{10}~(IF); ~~~ b = B_{+-}=-B_{-+}~(FF) \\
\label{15e}
\alpha,\beta &=& 0,1 ~~~ and ~~~ \mu =0,1,i;~~~ i=2,3,....,25~~(IF) \\
\label{15f}
\alpha,\beta &=& +,- ~~~ and ~~~ \mu =+,-,i;~~~ i=2,3,....,25~~(LC/FF)
\label{15g}
\end{eqnarray}
\reseteqn
The theory described by $S^{I}$ is GI and is indeed seen to posses three local gauge symmetries given by the two dimensional WSRI and the WI defined by \cite{1,2}:
\alpheqn
\begin{eqnarray}
X^\mu &\longrightarrow & \tilde X^\mu = [X^\mu + \delta X^\mu] \quad ; \quad
\delta X^\mu = [\zeta^\alpha (\partial_\alpha X^\mu)] \\
\label{16a}
h^{\alpha\beta} &\longrightarrow& \tilde{h}^{\alpha\beta} = [h^{\alpha\beta} + \delta h^{\alpha\beta}] \quad ; \quad
\delta h^{\alpha\beta} = \left [\zeta^{\gamma} \partial_\gamma h^{\alpha\beta} - \partial_\gamma \zeta^\alpha h^{\gamma\beta} - \partial_\gamma \zeta^\beta h^{\alpha\gamma} \right ] \\
\label{16b}
h_{\alpha\beta} &\longrightarrow & [\Omega h_{\alpha\beta}]  \\
\label{16c}
B_{\alpha\beta} &\longrightarrow & {\mathop{B}^{\sim}}_{\alpha\beta} = [B_{\alpha\beta} + \delta B_{\alpha\beta}] \quad ; \quad 
\delta B_{\alpha\beta} =  [\zeta^{\alpha} \partial_{\alpha} B_{\alpha \beta}] \label{16d}
\end{eqnarray}
\reseteqn
In the FF/LCQ the action $S^{I}$ reads\cite{2}:
\alpheqn
\begin{eqnarray}
S_{LC} &=& \int {\cal L}_{LC}~~ d\sigma^+ d\sigma^- \\
\label{17a}
{\cal L}_{LC} &=& \left[\frac{- \lambda T}{2} \right] \left[ (\partial_+ X^+) (\partial_- X^-) + (\partial_+ X^-)(\partial_- X^+) + (\partial_+ X^i) (\partial_- X^i) - \Lambda T b \right] 
\label{17b}
\end{eqnarray} 
\reseteqn
and the theory is seen to possess twenty seven constraints:
\alpheqn
\begin{eqnarray}
\chi_1 &=& [P^+ + (\frac {\lambda T}{2}) (\partial_-X^+) ] \approx 0 \quad ; \quad \chi_2 = [P^- + (\frac {\lambda T}{2})(\partial_-X^-) ] \approx 0 \\
\label{18a}
\chi_3 &=& [P_i + (\frac {\lambda T}{2}) (\partial_-X^i) ] \approx 0 \quad ; \quad \chi_4 = \Pi_b  \approx 0 
\label{18b}
\end{eqnarray}
\reseteqn
Where $P^+,~ P^-$, $P_i$ and $\Pi_b$ are the canonical momenta conjugate respectively to $X^-,~ X^+$, $X_i$ and~ b. These constraints form a set of first-class constraints and the theory could be quantized under appropriate guage-fixing. The action $S^{I}$ in the IF reads\cite{2}:
\begin{eqnarray}
S_{IF} = \int {\cal L}_{IF}~~ d\tau d\sigma \quad ; \quad  {\cal L}_{IF} = \left [ (- \lambda T/2)  [ (X{'})^2 - (\dot X)^2 ]-\Lambda T b \right ]  
\label{19}
\end{eqnarray}
The canonical momenta conjugate respectively to $X_\mu$ and $b(\equiv B_{01}=-B_{10})$ are denoted by  $P^\mu$ and $\Pi_b$. The theory described by $S_{IF}$ is seen to possess only one primary constraint: \quad $\Psi_1 = \Pi_b \approx 0 $. The total Hamiltonian density of the theory could be obtained by incorporating the primary constraints of the theory $\Psi_1$ in the canonical Hamiltonian density ${\cal H}^c_{IF}$ with the help of Lagrange multiplier field $u{(\tau,\sigma)}$ (which is treated as dynamical):
\alpheqn
\begin{eqnarray}
{\cal H}^T_{IF} &=& [ {\cal H}^c_{IF} + u \Psi_1 ] \\
\label{20a}
&=& \left [ (1/ (2\lambda T))P^{\mu}P_\mu + (\lambda T/2)(X^{'})^2 + \Lambda T b + u \Pi_b \right ]
\label{20b}
\end{eqnarray}
\reseteqn
Demanding that the primary constraint $\Psi_1$ be preserved in the course of time one does not get any further constraints. The theory is thus seen to posses only one constraint $\Psi_1$. Also, because the constraint $\Psi_1$ is first-class, the theory described by $S_{IF}$ is GI and it could be quantized under appropriate gauge-fixing.  To study the Hamiltonian and path integral formulations of the theory described by $S_{IF}$ under gauge-fixing, we could choose e.g., the gauge: ~ $ \zeta $ = $ b \approx 0 $. Corresponding to this choice of gauge the total set of constraints of the theory under which the quantization of the theory could e.g., be studied becomes:
\begin{eqnarray}
\Psi_1 &=& \Pi_b \approx 0  \quad ; \quad
\Psi_2 = \zeta = b \approx 0
\label{21}
\end{eqnarray}
The matrix  $ M_{ij} (:=\{\Psi_i , \Psi_j \}_{PB})$ of the Poisson brackets of the constraints $ \Psi_i $ is seen to be nonsingular implying that the corresponding set of constraints $\Psi_i$ is a set of second-class constraints. The determinant of the matrix $M_{ij}$ is given by:
\begin{eqnarray}
[ \|\det (M_{ij})\|]^{1/2} &=& \delta (\sigma-\sigma') 
\label{22}
\end{eqnarray}
Now, following the standard Dirac quantization procedure in the Hamiltonian formulation \cite{2,3,4}, the nonvanishing EWST commutation relations of the theory described by the action $S_{IF}$ under the gauge: ~  $\zeta = b \approx 0$~~ are obtained as \cite{2}:
\begin{eqnarray}
[X^\mu(\sigma,\tau), P^{\nu}(\sigma',\tau)] = (i)~\eta^{\mu \nu}~\delta(\sigma-\sigma') 
\label{23}
\end{eqnarray}
In the path integral formulation, the transition to the quantum theory, is, however, made by writing the vacuum to vacuum transition amplitude called the generating functional $Z [J_i]$ of the theory under the gauge $\zeta = b \approx 0$,   in the presence of external sources $J_i$ as \cite{2} :
\begin{equation}
Z [J_i] := \int[d\mu] \exp \left [i \int d^2 \sigma [ P^\mu(\partial_\tau X_\mu) + \Pi_b (\partial_\tau b) + p_u(\partial_\tau u) - {\cal H}^T_{IF}  + J_i\Phi^i] \right ]
\label{24}
\end{equation}
Where $ p_u$  is the momentum canonically conjugate to  u.  Also the phase space variables of the theory are $\Phi^i \equiv (X^\mu,b,u)$ with the corresponding respective canonical conjugate momenta $\Pi_i \equiv (P_\mu,\Pi_b,p_u)$. The functional measure $[d\mu]$ of the generating functional $Z [J_i]$ under the gauge: ~ $\zeta = b \approx 0 $ ~ is:
\begin{eqnarray}
[d\mu] &=& [\delta (\sigma-\sigma')] [dX^\mu][db][du] [dP_\mu][d\Pi_b] \nonumber \\
& & [dp_u] \cdot \delta [(\Pi_b) \approx 0] \cdot \delta [(b) \approx 0].
\label{25}
\end{eqnarray}

It is important to point out that the 2-form gauge field $B_{\alpha\beta}$  is seen here to behave like a Wess-Zumino (WZ) field and the term involving this field is seen to behave like a WZ term for the CGFPD1BA. Also the gauge $\zeta = b \approx 0$, translates the GI system $S^{I}$ into the corresponding GNI system  $S^{N}$ (for the theory in the IF as well as in the FF/LCQ) \cite{2}, and the physics content of the two systems remains the same.

About the BC´s, it is important to mention here that in principle, there are two two different ways to take them in to account: either by imposing them directly in the usual way for the open and closed strings separately in an appropriate manner \cite{1} or by considering them as the Dirac primary constraints \cite {6} and study them accordingly.

Further, in the usual Hamiltonian and path integral formulations of a GI theory under some GFC's, one necessarily destroys the gauge invariance of the theory by fixing the gauge (which converts a set of first class constraints into a set of second-class constraints, implying a breaking of gauge invariance under the gauge fixing).  To achieve the quantization of a GI theory such that the gauge invariance of the theory is maintained even under gauge fixing, one goes to a more generalized procedure called the Becchi-Rouet-Stora and Tyutin (BRST) BRST formulation \cite{3}. In the BRST formulation of a GI theory, the theory is rewritten as a quantum system that possesses a generalized gauge invariance called the BRST symmetry.  For this one enlarges the Hilbert space of the gauge invariant theory and replaces the notion of the gauge transformation, which shifts operators by c-number functions, by a BRST transformation, which mixes operators having different statistics.  In view of this, one introduces new anti-commuting variables called the Faddeev-Popov ghost and anti-ghost fields, which are Grassmann numbers on the classical level and operators in the quantized theory, and the commuting variables called the Nakanishi-Lautrup fields \cite{3}. In the BRST formulation, one thus embeds a GI theory into a BRST-invariant system, and the quantum Hamiltonian of the system (which includes the gauge fixing contribution) commutes with the BRST charge operator $Q$ as well as with the anti-BRST charge operator $\overline{Q}$. The new symmetry of the quantum system (the BRST symmetry) that replaces the gauge invariance is maintained even under the gauge fixing and hence projecting any state onto the sector of BRST and anti-BRST invariant state yields a theory which is isomorphic to the original GI theory. The technical details of the calculations are omitted here for the sake of brevity. 

\section{Acknowledgements}

It is a matter of great pleasure to thank the organizers of THEP-I, and in particular to thank Dr. Aalok Mishra and the distinguished faculty members of IIT Roorkee as well as the team of  younger people who provided a great stimulating atmosphere during the workshop.

\end{document}